\documentclass[final]{IEEEtran}
\pdfoutput=1
\usepackage{amsthm,amssymb,graphicx,multirow,amsmath,color,amsfonts}%,ulem}

\usepackage{subcaption}
\usepackage{caption}
\usepackage[update,prepend]{epstopdf}
\usepackage[noadjust]{cite}
\usepackage[latin1]{inputenc}
\usepackage{tikz}
\usepackage{bbm} % for \mathbbm{1}
\usepackage{pdfpages}
\usepackage{multirow}
\usepackage{comment}

\usepackage{algorithm}
\usepackage{algpseudocode}
\usepackage{url}
\def\BSTATE{\STATE\hskip-\ALG@thistlm}
\makeatother

\def\nb0{{\mathbf{0}}}
\def\nb1{{\mathbf{1}}}

\newtheorem{theorem}{Theorem}

\def\argmin{\operatorname{arg~min}}

\def\P{\mathbb{P}}

\def\g{\left.\right|}

\allowdisplaybreaks % Allows breaking of eqnarray over multiple pages (avoids unnecessary blanks in the document before eqnarray)

\usepackage{setspace}	% Remove in double column version. Also search for \setstretch in the body of the paper and comment these commands for double column

\begin{document}
%\pagenumbering{gobble}
\graphicspath{{./Figures/}}
\title{
Deep Reinforcement Learning for Minimizing Age-of-Information in UAV-assisted Networks}
\author{
Mohamed A. Abd-Elmagid\IEEEauthorrefmark{1}, Aidin Ferdowsi\IEEEauthorrefmark{1}, Harpreet S. Dhillon, and Walid Saad
\thanks{The authors are with Wireless@VT, Department of ECE, Virginia Tech, Blacksburg, VA (Email: \{maelaziz,aidin,hdhillon,walids\}@vt.edu). The support of the U.S. NSF (Grant CNS-1814477) is gratefully acknowledged.

\IEEEauthorrefmark{1} The first two authors have equally contributed to this paper and are listed  according to the alphabetical order. }
\vspace{-5mm}
}
%This research was supported by the U.S. National Science Foundation under Grants --- and ---.

\maketitle

\begin{abstract}
Unmanned aerial vehicles (UAVs) are expected to be a key component of the next-generation wireless systems. Due to their deployment flexibility, UAVs are being considered as an efficient solution for collecting information data from ground nodes and transmitting it wirelessly to the network. In this paper, a UAV-assisted wireless network is studied, in which energy-constrained ground nodes are deployed to observe different physical processes. In this network, a UAV that has a time constraint for its operation due to its limited battery, moves towards the ground nodes to receive status update packets about their observed processes. The flight trajectory of the UAV and scheduling of status update packets are jointly optimized with the objective of achieving the minimum weighted sum for the age-of-information (AoI) values of different processes at the UAV, referred to as \emph{weighted sum-AoI}. The problem is modeled as a finite-horizon Markov decision process (MDP) with finite state and action spaces. Since the state space is extremely large, a deep reinforcement learning (RL) algorithm is proposed to obtain the optimal policy that minimizes the weighted sum-AoI, referred to as the \emph{age-optimal policy}. Several simulation scenarios are considered to showcase the convergence of the proposed deep RL algorithm. Moreover, the results also demonstrate that the proposed deep RL approach can significantly improve the achievable sum-AoI per process compared to the baseline policies, such as the distance-based and random walk policies. The impact of various system design parameters on the optimal achievable sum-AoI per process is also shown through extensive simulations.
\end{abstract}\vspace{-3mm}
\section{Introduction} \label{sec:intro}
Owing to their deployment flexibility, unmanned aerial vehicles (UAVs) are expected to be a key component of future wireless networks. The use of UAVs as flying base stations, that collect/transmit information from/to ground nodes (e.g., users, sensors or Internet of Things (IoT) devices), has recently attracted significant attention \cite{mozaffari2019tutorial,challita2019machine,azari2016joint,bor2016efficient,chetlur2017downlink,
alzenad20173,zeng2016throughput,li2018placement,xie2018throughput,monwar2018optimized}. Meanwhile, introducing UAVs into wireless networks leads to many challenges such as optimal deployment, flight trajectory design, and energy efficiency. So far, these challenges have mainly been addressed in the literature with the objective of either maximizing network coverage and rate or minimizing delay. In contrast, the quality-of-service (QoS) for many real-time applications, e.g., human safety applications, is restricted by the freshness of information collected by the UAV from the ground nodes \cite{abd2018role}. Therefore, in order to preserve the freshness of information status at the UAV, it is important to design its flight trajectory as well as carefully schedule information transmissions from the ground nodes.

 {\it Related works.} We employ the concept of age-of-information (AoI) to quantify the freshness of information at the UAV. First introduced in \cite{kaul2012real}, AoI is defined as the time elapsed since the latest received status update packet at a destination node was generated at the source node. For a simple queueing-theoretic model, the authors of \cite{kaul2012real} characterized the average AoI. Then, the average AoI and some other age-related metrics were investigated in the literature for variations of the queueing model considered in \cite{kaul2012real} (refer to \cite{kosta2017age_mono} for a comprehensive survey). Another line of research \cite{ABedewy2016,sun2017update,8648525,
113882,AbdElmagid2019Globecom_a,talak2018optimizing,zhou2018joint,abdel2018ultra,abd2018average,8406973} employed AoI as a performance metric for different communication systems that deal with time critical information. The main focus of these works was on applying tools from optimization theory to characterize age-optimal transmission policies. Note that the destination node was commonly assumed to be a static node in \cite{kaul2012real,kosta2017age_mono,sun2017update,ABedewy2016,8648525,
113882,AbdElmagid2019Globecom_a,talak2018optimizing,zhou2018joint,abdel2018ultra,abd2018average,8406973}. 
%we study the scenario in which a mobile destination node, represented by the UAV, moves towards the ground nodes to collect status update packets about their observed processes. For this setup, we jointly design the UAV's flight trajectory as well as scheduling of update packet transmissions to achieve the minimum weighted sum-AoI.
More recently, in \cite{abd2018average} and \cite{8406973}, the authors considered the optimization of AoI in UAV-assisted wireless networks. However, the analyses in these works were limited to scenarios where UAVs acted as relay nodes and are hence not broadly applicable. Furthermore, these works did not take into account the optimal scheduling of update packet transmissions from different nodes while optimizing the UAV's flight trajectory. 

{\it Contributions.} The main contribution of this paper is a novel deep reinforcement learning (RL) framework for optimizing the UAV's flight trajectory as well as scheduling status update packets from ground nodes with the objective of minimizing the weighted sum-AoI. In particular, we study a UAV-assisted wireless network, in which a UAV moves towards the ground nodes to collect status update packets about their observed processes. For this system setup, we formulate a weighted sum-AoI minimization problem in which the UAV's flight trajectory as well as scheduling of update packet transmissions are jointly optimized. To obtain the age-optimal policy, the problem is first modeled as a finite-horizon Markov decision process (MDP) with finite state and action spaces. Due to the extreme curse of dimensionality in the state space, the use of a finite-horizon dynamic programming (DP) algorithm is computationally impractical. To overcome this challenge, we propose a deep RL algorithm. After showing the convergence of our proposed algorithm, we numerically demonstrate its superiority over two baseline policies, namely, the distance-based and random walk policies, in terms of the achievable sum-AoI per process. Several key system design insights are also provided through extensive numerical results. To the best of our knowledge, this work is the first to apply tools from deep RL to characterize the age-optimal policy.
%drawn from our numerical results. For instance, the achievable sum-AoI per process by the proposed algorithm is monotonically increasing (monotonically decreasing) with the time constraint of the UAV and spatial density of the ground nodes (the battery sizes of the ground nodes).
 %while jointly optimizing the UAV's flight trajectory as well as scheduling of status update packets. 
%%
%Particularly, under the distance-based policy, the UAV receives an update packet from its closest ground node while moving towards the node with the largest AoI. On the other hand, as per the random walk policy, the UAV's movement as well as scheduling of update packet transmissions are decided randomly. 
\vspace{-3mm}
\section{System Model}\label{sec:Model}
\subsection{Network Model}
Consider a wireless network in which a set $\mathcal{M}$ of $M$ ground nodes are deployed to observe potentially different physical processes of a certain geographical region. Uplink transmissions are considered, where a UAV collects status update packets from the nodes while seeking to maintain freshness of its information status about their observed processes during the time of its operation. We assume a discrete-time system in which time is divided into slots of unit length (without loss of generality) such that each slot $n$ corresponds to the time duration $[n - 1, n)$. Each ground node $m$ has a battery with finite capacity $E_{{\rm max},m}$, which is divided into a finite number of energy quanta $e_{{\rm max},m}$ such that the amount of energy contained in each energy quantum is $E_{{\rm max},m} / e_{{\rm max},m}$. Let $e_{m}(n) \in \mathcal{E}_{m}\hspace{-0.5mm} \triangleq\hspace{-0.5mm} \left\{0,1,\cdots,e_{m,{\rm max}} \right\}$ denote the battery level at device $m$ at the beginning of slot $n$.

\begin{figure}[t!]
\centering    
\includegraphics[width=0.75\columnwidth]{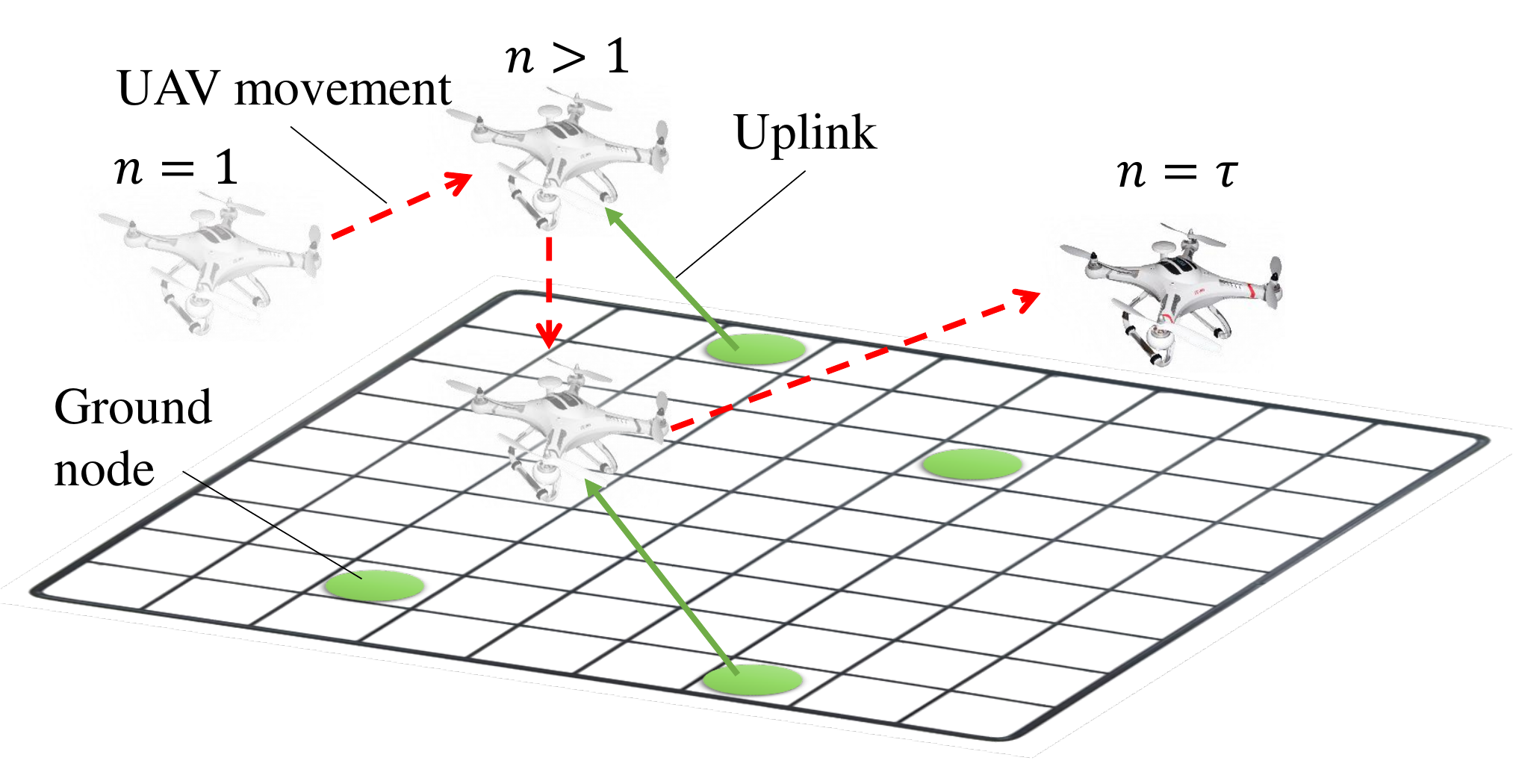}
    \caption{\small An illustration of our system model.}
    \label{fig:system model}
\end{figure}

As shown in Fig. \ref{fig:system model}, the geographical region of interest is partitioned into cells of equal areas where we denote by $L_{{\rm c},i} = (x_{{\rm c},i},y_{{\rm c},i}) \in \mathcal{C}$ the location of the center of cell $i$, and $\mathcal{C}$ is the set containing the locations of centers for different cells. Let $x_{\rm s}$ and $y_{\rm s}$ be the horizontal and vertical spacing distances between the centers of any two adjacent cells, respectively. The UAV is assumed to fly at a fixed height $h$ such that the projection of its flight trajectory on the ground at time slot $n$ is denoted by $L_{u}(n) \in \mathcal{C}$. In other words, we will discretize the trajectory of the UAV such that its location is mapped to a discrete value $L_u(n)$ during time slot $n$. In practice, the UAV can only operate for a finite time interval due to its battery limitations and the need for recharging. We model this fact by having a time constraint of $\tau$ seconds during which the UAV flies from an initial location $L_{u}^{\rm i}$ to a final location $L_{u}^{\rm f}$ where it can be recharged to continue its operation. Note that $L_{u}^{\rm i}$ and $L_{u}^{\rm f}$ are the center locations of the initial and final cells, respectively, along the UAV's flight trajectory. Therefore, the UAV's flight trajectory is approximated by the sequence $\left \{ L_{u}^{\rm i},\cdots,L_{u}(n),\cdots,L_{u}^{\rm f} \right \}$. Similar to \cite{zeng2016throughput,li2018placement,xie2018throughput}, the channels between the UAV and ground nodes are assumed to be dominated by the line-of-sight (LoS) links. Therefore, at time slot $n$, the channel power gain between the UAV and ground node $m$ will be:
\begin{align}\label{eq:channel}
g_{u,m}(n) = \beta_0 d_{u,m}^{- 2} = \frac{\beta_0}{h^2 + \lVert L_{u}(n) - L_{m}\rVert^2},\; m \in \mathcal{M},
\end{align} 
where $d_{u,m}$ is the distance between the UAV and node $m$, $L_{m}$ is the location of node $m$, and $\beta_0$ is the channel gain at a reference distance of 1 meter.

 The AoI of an arbitrary physical process is defined as the time elapsed since the most recently received update packet at the UAV was generated at the ground node observing this process. Denote by $A_{m}(n) \in \mathcal{A}_{m} \triangleq \left \{1,2,\cdots,A_{{\rm max},m}\right \}$ the AoI at the UAV for the process observed by node $m$ at the beginning of time slot $n$, where $A_{{\rm max},m}$ is the maximum value that $A_{m}(n)$ can take and can be chosen to be arbitrary large.
%  \cite{abd2018role,zhou2018joint}. 
% We use the concept of AoI to quantify the freshness of information status at the UAV.
\subsection{State and Action Spaces}
 The state $s_{m}(n)$ of a ground node $m$ at time slot $n$ is characterized by its battery level and the AoI of its observed process at the UAV at the beginning of slot $n$, i.e., $s_{m}(n) \triangleq \left(e_{m}(n),A_{m}(n) \right)\in \mathcal{E}_{m} \times \mathcal{A}_{m}$. On the other hand, the state of the UAV $s_{u}(n)$ at slot $n$ is captured by its location, and the difference between the remaining time and the required time to reach its final location which is denoted by $t_{u}(n) \in \mathcal{T} \triangleq \{0,1,\cdots,\tau\}$, i.e., $s_{u}(n) \triangleq \left(L_{u}(n), t_{u}(n)\right) \in \mathcal{C} \times \mathcal{T}$. Hence, the system state at slot $n$ can be expressed as $s(n) = \left(\left \{s_{m}(n)\right\}_{m \in \mathcal{M}}, s_{u}(n)\right) \in \mathcal{S} \triangleq \prod_{m \in \mathcal{M}}{(\mathcal{E}_{m} \times \mathcal{A}_{m})} \times \mathcal{C} \times \mathcal{T}$, where $\mathcal{S}$ is the state space of the system.

We assume that UAV's maximum allowable speed limits its movement in each slot to one of the adjacent cells of its current cell. Hence, in each time slot, the UAV either decides to remain at its location over the duration of the next slot or move to one of its adjacent cells. Let $v(n) \in \mathcal{V} \triangleq \{N,S,E,W,I\}$ be the movement action of the UAV at slot $n$, where $N,S,W$ and $E$ denote the north, south, west and east directions, respectively, and $I$ indicates that the UAV will remain at its location in the next slot. Hence, the dynamics of the UAV's location will be:
\begin{align}\label{eq:UAV_loc_evol}
L_{u}(n + 1) = \begin{cases}
\begin{aligned}
&L_{u}(n) + (0,y_{\rm s}),\; &&\text{if}\; v(n) = N,\\
&L_{u}(n) - (0,y_{\rm s}),\; &&\text{if}\; v(n) = S,\\
&L_{u}(n) + (x_{\rm s},0),\; &&\text{if}\; v(n) = E,\\
&L_{u}(n) - (x_{\rm s},0),\; &&\text{if}\; v(n) = W,\\
&L_{u}(n),\; &&\text{otherwise}.
\end{aligned}
\end{cases}
\end{align}

When the UAV is located at one of the boundary cells in time slot $n$ and $v(n)$ will cause its location to be outside the considered region in slot $n + 1$, it will remain at its current location in slot $n + 1$. Meanwhile, at each slot, the UAV may choose one of the ground nodes from which it receives an update packet about its observed process. Let $w(n) \in \mathcal{W} = \{0,1,\cdots,M\}$ denote the scheduling action for update packet transmission at slot $n$, where $w(n) = m$ means that node $m$ is scheduled to transmit an update packet at slot $n$, and $w(n) = 0$ indicates that no update packet transmission occurs at slot $n$. Hence, the system action at slot $n$ is $a(n) = \left(v(n), w(n)\right) \in \mathcal{A} \triangleq \mathcal{V} \times \mathcal{W}$, where $\mathcal{A}$ is the action space of the system.

By letting $S, B$, and $\sigma^2$ be the size of an update packet, channel bandwidth, and noise power at the UAV, respectively, the integer number of energy quanta required to transmit an update packet from node $m$ is given by:
\begin{align}\label{eq:energy_dis}
e_{m}^{\rm T}(n) = \left \lceil \frac{e_{{\rm max},m}}{E_{{\rm max},m}} E_{m}^{\rm T}(n)\right \rceil,
\end{align}
where, according to Shannon's formula and recalling that the slot length is unity, $E_{m}^{\rm T}(n)$ can be expressed as:
 \begin{align}\label{eq:energy_con}
E_{m}^{\rm T}(n) = \frac{\sigma^2}{g_{u,m}(n)}\left(2^{S/B} - 1\right).
\end{align}

 Clearly, when node $m$ is scheduled to transmit an update packet at slot $n$, its current battery level $e_{m}(n)$ should be at least equal to $e_{m}^{\rm T}(n)$. The ceiling was used in (\ref{eq:energy_dis}) to obtain a lower bound on the performance of the continuous system (when the energy in the battery is expressed by a continuous variable). On the other hand, if the floor operator replaces the ceiling one in the definition of $e_{m}^{\rm T}(n)$, an upper bound on the performance of the continuous system is obtained. Therefore, the evolution of the battery level at node $m$ is given by
\begin{align}\label{eq:batt_evol}
e_{m}(n + 1) = \begin{cases}
\begin{aligned}
&e_{m}(n) - e_{m}^{\rm T}(n),\; &&\text{if}\; w(n) = m,\\
&e_{m}(n),\; &&\text{otherwise}.
\end{aligned}
\end{cases}
\end{align}

A {\it generate-at-will policy} is employed such that whenever node $m$ is chosen to transmit an update packet at a certain time slot, it generates that update packet at the beginning of that time slot \cite{sun2017update,8648525}. Therefore, when $w(n) = m$, the AoI of its observed process reduces to one; otherwise, the AoI value increases by one. Hence, the AoI dynamics for the process observed by node $m$ can be expressed as 
\begin{align}\label{eq:AoI_evol}
A_{m}(n + 1) = \begin{cases}
\begin{aligned}
&1,\; &&\text{if}\; w(n) = m,\\
&{\rm min}\left \{A_{{\rm max},m}, A_m(n)+1 \right \},\; &&\text{otherwise}.
\end{aligned}
\end{cases}
\end{align}
\section{Deep Reinforcement Learning for Weighted Sum-AoI Minimization}\label{sec:DQN}
\subsection{Problem Formulation}
Our goal is to characterize the age-optimal policy which determines the actions decided at different states of the system over a finite horizon of length $\tau$. The objective of this age-optimal policy is to minimize the weighted sum-AoI. Formally, a policy $\pi = \{\pi_n\}, n = 1,2,\cdots, \tau,$ is a sequence of probability measures over the state space $\mathcal{S}$. Let $s^{(n)} = \left \{s(1),a(1),\cdots,s(n - 1),a(n - 1),s(n)\right \}$ denote the sequence of actions and states up to the state of the system at slot $n$. Conditioned on $s^{(n)}$, the probability measure $\pi_{n}$ determines the probability of taking action $a(n)$, i.e., $\P\left(a(n) \g s^{(n)}\right)$. In addition, the policy $\pi$ is called stationary when $\P\left(a(n) \g s^{(n)}\right) = \P\left(a(n) \g s(n)\right), \forall n,$ and is said to be deterministic when $\P\left(a(n) \g s^{(n)}\right) = 1$ for some $a(n) \in \mathcal{A}(s(n))$, where $\mathcal{A}(s(n))$ represents the set of possible actions at state $s(n)$. Given a policy $\pi$, the total expected cost of the system, over the finite horizon of interest starting from an initial state $s(1)$, can be expressed as
\begin{align}\label{average_AoI}
G^{\pi}\left(s^{(\tau)}\right) \triangleq \sum_{n = 1}^{\tau} \sum_{m = 1}^{M} {\lambda_{m} \mathbb{E}\left[{A_m(n)} \g s(1)\right]}, 
\end{align}
where $\lambda_{m}$ is the importance weight of the process observed by node $m$ and the expectation is taken with respect to the policy. Our goal is to obtain the optimal policy $\pi^\star$ that satisfies 
\begin{align}\label{Optim_policy}
\pi^\star = {\rm arg}\; \underset{\pi}{\rm min} \;G^{\pi}\left(s^{(\tau)}\right).
\end{align}

Next, we derive the maximum and minimum total expected costs for identical ground nodes which have equal numbers of energy quanta, importance weights, and maximum AoI values.
\begin{theorem}
The maximum and minimum total expected costs of the system, for a case with identical ground nodes, $A_{\max,m} = A_{\max} \geq M$, and $A_{m}(1) = 1, \forall m$, are given by:
\begin{align}
\min_{\pi} G^{\pi} &= \frac{(2M+1)(M-1)}{4}-\sum_{n=1}^{M-1} \frac{n^2}{2M}\nonumber\\& +\frac{ (\tau - (M+1))(M+1)}{2},\label{eq:min}\\
\max_{\pi} G^{\pi} &= \frac{A_{\max}(A_{\max}-1)}{2} + (\tau - (A_{\max}-1) )A_{\max}.\label{eq:max}
\end{align}\normalsize
\end{theorem}
\begin{IEEEproof}
The minimum total expected cost is reached when the UAV can receive an update packet from the ground node with maximum current AoI value at every time slot. In this case, we have:
\begin{align}
\sum_{m=1}^M\lambda_m A_m(n) = 
\begin{cases}
1/M\left[nM - \frac{n(n-1)}{2}\right], & n<M\\
\frac{(M+1)}{2},&n\geq M.
\end{cases}
\end{align}
By summing this value over all time slots, we obtain \eqref{eq:min}.
The maximum total expected cost is reached when the UAV cannot receive update packets over all time slots. In this case, we have:
\begin{align}\label{eq:maxcases}
\sum_{m=1}^M\lambda_m A_m(n) =
\begin{cases}
n, & n<A_{\max},\\
A_{\max}, & n\geq A_{\max}.
\end{cases}
\end{align}
Thus, by summing \eqref{eq:maxcases} over all time slots, we get \eqref{eq:max}.
\end{IEEEproof}

Owing to the nature of evolution of the system state parameters, represented by (\ref{eq:UAV_loc_evol}), (\ref{eq:batt_evol}), and (\ref{eq:AoI_evol}), the problem can be modeled as a finite-horizon MDP with finite state and action spaces. However, due to the curse of extremely high dimensionality in $\mathcal{S}$, it is computationally infeasible to obtain $\pi^\star$ using the standard finite-horizon DP algorithm \cite{powell2007approximate}. Motivated by this, we propose a deep RL algorithm for solving (\ref{Optim_policy}) in the next subsection. Deep RL is suitable for this problem since it can reduce the dimensionality of the large state space while learning the optimal policy at the same time \cite{mnih2015human}.\vspace{-3mm}
\subsection{Deep Reinforcement Learning Algorithm}
\begin{figure}[t!]
\centering
    \includegraphics[width=0.75\columnwidth]{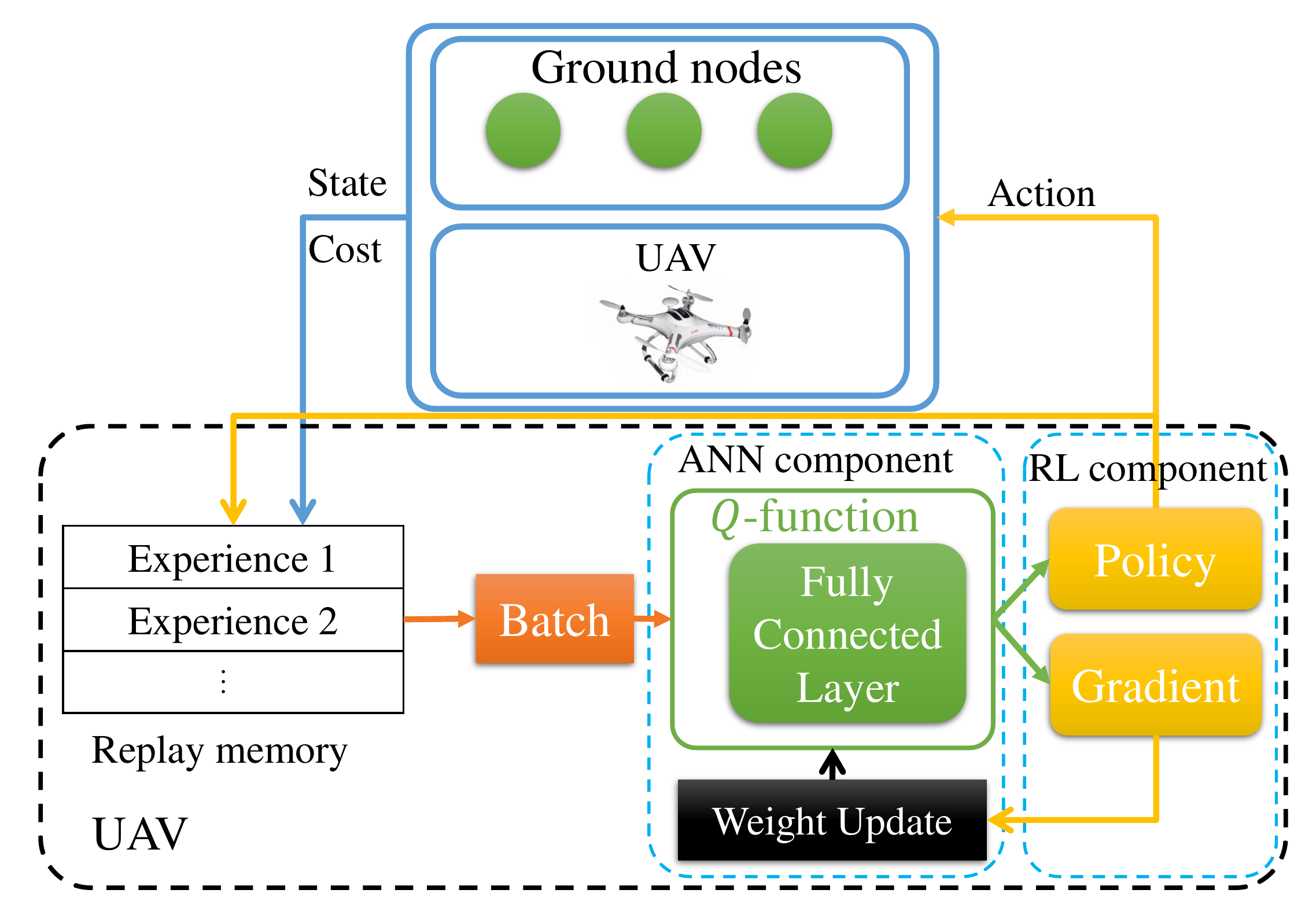}
    \caption{\small The deep RL architecture.}
    \label{fig:deepRL}
\end{figure}

The proposed deep RL algorithm has two components:
(i) an artificial neural network (ANN), that reduces the dimension of the state space by extracting its useful features and (ii) an RL component, which is used to find the best policy based on the ANN's extracted features, as shown in Fig. \ref{fig:deepRL}.

To derive the policy that minimizes the total expected cost of the system, we use a $Q$-learning algorithm \cite{powell2007approximate}. In this algorithm, we define a state-action value function $Q^\pi(s(n),a(n))$ which is the expected cost of the system starting at state $s(n)$, performing action $a(n)$ and following policy $\pi$. In $Q$-learning algorithm, we try to estimate the $Q$-function using any policy that minimizes the future cost. To this end, we use the so-called Bellman update rule as follows:
\begin{align}\label{eq:Bellman}
&Q_{n+1}\left(s(n),a(n)\right) = Q_{n}\left(s(n),a(n)\right) + \beta \Big(c(n)\nonumber \\
&+ \gamma \min_{\alpha} Q_{n}\left(s(n+1),\alpha\right) - Q_{n}\left(s(n),a(n)\right) \Big),
\end{align}\normalsize
where $c(n) = \sum_{m=1}^M {\lambda_m A_m(n)}$ represents the instantaneous cost at slot $n$, $\beta$ is the learning rate, and $\gamma$ is a discount factor. The discount factor can be set to a value between 0 and 1 if the UAV's task is \emph{continuing} which means the task will never end, thus the current cost has higher value than the unknown future cost. However, in our case, we have two terminal cases: 1) when the UAV reaches the final cell and 2) when the required time to go to the final cell is less than the remaining time slots due to the limited available energy of the UAV. Therefore, our problem is \emph{episodic}, and hence we set $\gamma = 1$.
% because all time slots have equal values for the UAV.

Since, using \eqref{eq:Bellman}, the UAV always has an estimate of the $Q$-function, it can \emph{exploit} the learning by taking the action that minimizes the cost. However, when learning starts, the UAV does not have confidence on the estimated value of the $Q$-function since it may not have visited some of the state-action pairs. Thus, the UAV has to \emph{explore} the environment (all state-action pairs) with some degree. To this end, an $\epsilon$-greedy approach is used where $\epsilon$ is the probability of exploring the environment at the current state \cite{mnih2015human}, i.e., taking a random action with some probability. One can reduce the value of $\epsilon$ to $0$ as the learning goes on to insure that the UAV chooses the optimal action rather than explore the environment.

The iterative method in \eqref{eq:Bellman} can be applied efficiently for the case in which the number of states is small. However, in our problem, the state space is extremely large which makes such an iterative approach impractical, since it requires a large memory and will have a slow convergence rate. Also, this approach cannot be generalized to unobserved states, since the UAV must visit every state and take every action to update every state-action pair \cite{powell2007approximate}. Thus, we employ ANNs which are very effective at extracting features from data points and summarizing them to smaller dimensions. We use a deep $Q$ network approach \cite{mnih2015human} in which the learning steps are the same as in $Q$-learning, however, the $Q$-function is approximated using an ANN $Q(s,a|\boldsymbol{\theta})$, where $\boldsymbol{\theta}$ is the vector containing the weights of the ANN. In particular, a fully connected (FC) layer, as in \cite{mnih2015human}, is used to extract abstraction of the state space. In the FC, every artificial node of a layer is connected to every artificial node of the next layer via the weight vector $\boldsymbol{\theta}$. The goal is to find the optimal values for $\boldsymbol{\theta}$ such that the ANN will be as close as possible to the optimal $Q$-function. To this end, we define a loss function for any set of $\left(s(n),a(n),c(n),s(n+1)\right)$, as follows:
\begin{align}
    L(\boldsymbol{\theta}_{k+1})& = \Big[c(n) + \gamma \min_{\alpha'} Q(s(n + 1),\alpha'|\boldsymbol{\theta}_{k}) \nonumber \\
    &- Q(s(n),a(n)|\boldsymbol{\theta}_{k+1})\Big]^2,
\end{align}
where subscript $k+1$ is the episode at which the weights are updated.
In addition, we use a \emph{replay memory} that saves the evaluation of the state, action, and cost of past \emph{experiences}, i.e., past state-actions pairs and their resulting costs. Then, after every episode, we sample a batch of $b$ past experiences from the replay memory and we find the gradient of the weights using this batch as follows:
\begin{align}\label{eq:gradient}
&\nabla_{\boldsymbol{\theta}_{k+1}} L(\boldsymbol{\theta}_{k+1}) = \Big[c(n) + \gamma \min_{\alpha'} Q(s(n + 1),\alpha'|\boldsymbol{\theta}_{k}) \nonumber \\
    &- Q(s(n),a(n)|\boldsymbol{\theta}_{k+1}) \Big]\times\nabla_{\boldsymbol{\theta}_{k+1}}Q(s(n),a(n)|\boldsymbol{\theta}_{k+1}).
\end{align}

Then, using this loss function, we train the weights of the ANN. It has been shown that using the batch method and replay memory improves the convergence of deep RL \cite{mnih2015human}.
Algorithm \ref{Algorithm:DeepRL} summarizes the proposed learning algorithm and Fig. \ref{fig:deepRL} shows the architecture of the deep RL algorithm.
\begin{algorithm}[t]
	\caption{Deep RL for weighted sum-AoI minimization}
	\begin{algorithmic}[1]\footnotesize 
		\State Initialize a \emph{replay memory} that stores the past experiences of the UAV and an ANN for $ Q $-function. Set $k=1$.
		\State \textbf{Repeat:}
		\State \quad Set $n=1$ and observe the initial state $ s(1) $.
		\State \quad \textbf{Repeat:}
		\State \quad \quad Select an action $ a $:
		\State \quad \quad \quad select a random action $a \in \mathcal{A}(s(n))$ with probability $ \varepsilon $ ,
		\State \quad \quad \quad otherwise select $ a = \argmin_{\alpha} Q(s(n),\alpha|\boldsymbol{\theta}_k) $.
		\State \quad \quad Perform action $ a $. 
		\State \quad \quad Observe the cost $ c(n) $ and the new state $ s(n+1) $.
		\State \quad \quad Store \emph{experience} $ \left\{s(n),a(n),c(n),s(n+1)\right\} $ in the replay \Statex \quad \quad memory.
		\State \quad \quad $ n = n+1 $
		\State \quad \textbf{Until} $ s(n+1) $ is a terminal state.
		\State \quad Sample $b$ random experiences $ \left\{\hspace{-0.5mm}\hat{s}(\eta), \hspace{-0.5mm}\hat{a}(\eta),\hspace{-0.5mm}\hat{c}(\eta),\hspace{-0.5mm}\hat{s}(\eta+1)\hspace{-0.5mm}\right\} $
		\Statex \quad from the replay memory.
		\State \quad Calculate the \emph{target} value $ t $:
		\State \quad \quad If the sampled experience is for $ \eta=1 $ then $ t=\hat{c}(\eta)  $,
		\State \quad \quad Otherwise $ t=\hat{c}(\eta) + \gamma \min_{\alpha'} Q(\hat{s}({\eta} + 1),\alpha'| \boldsymbol{\theta}_k) $.
		\State \quad Train the network $ Q $ using the gradient in \eqref{eq:gradient}.
		\State \quad $ k = k+1 $.
		\State \textbf{Until} convergence to a sum-AoI.
	\end{algorithmic}
	\label{Algorithm:DeepRL}
\end{algorithm}
\begin{figure*}[t!]
    \centering
    \begin{subfigure}[b]{0.25\textwidth}
    \centering
    \includegraphics[width=\columnwidth]{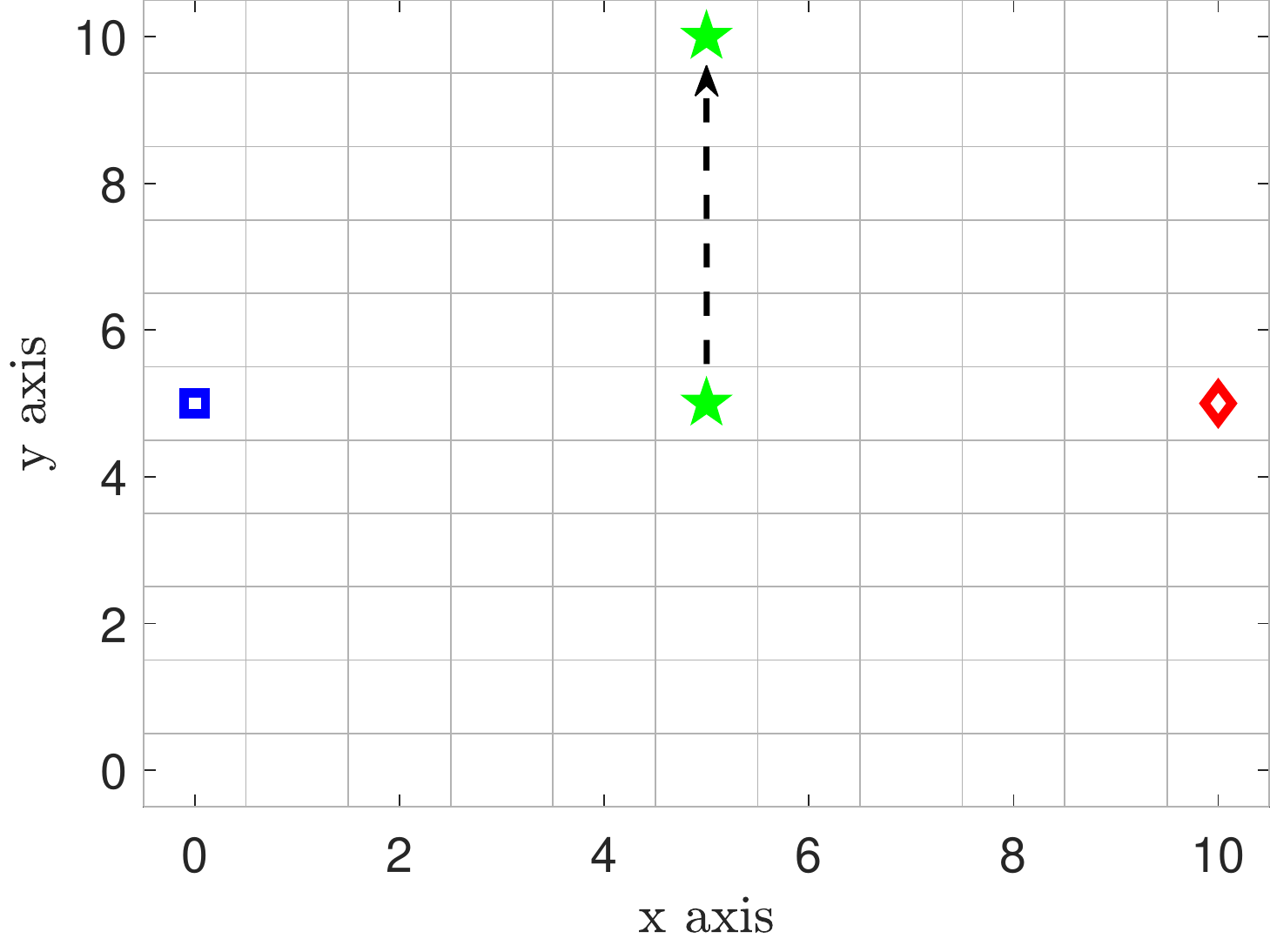}
    \caption{\small Setup for scenario 1.}
    \label{fig:setup_1}
    \end{subfigure}%
    ~
    \begin{subfigure}[b]{0.25\textwidth}
    \centering
    \includegraphics[width=\columnwidth]{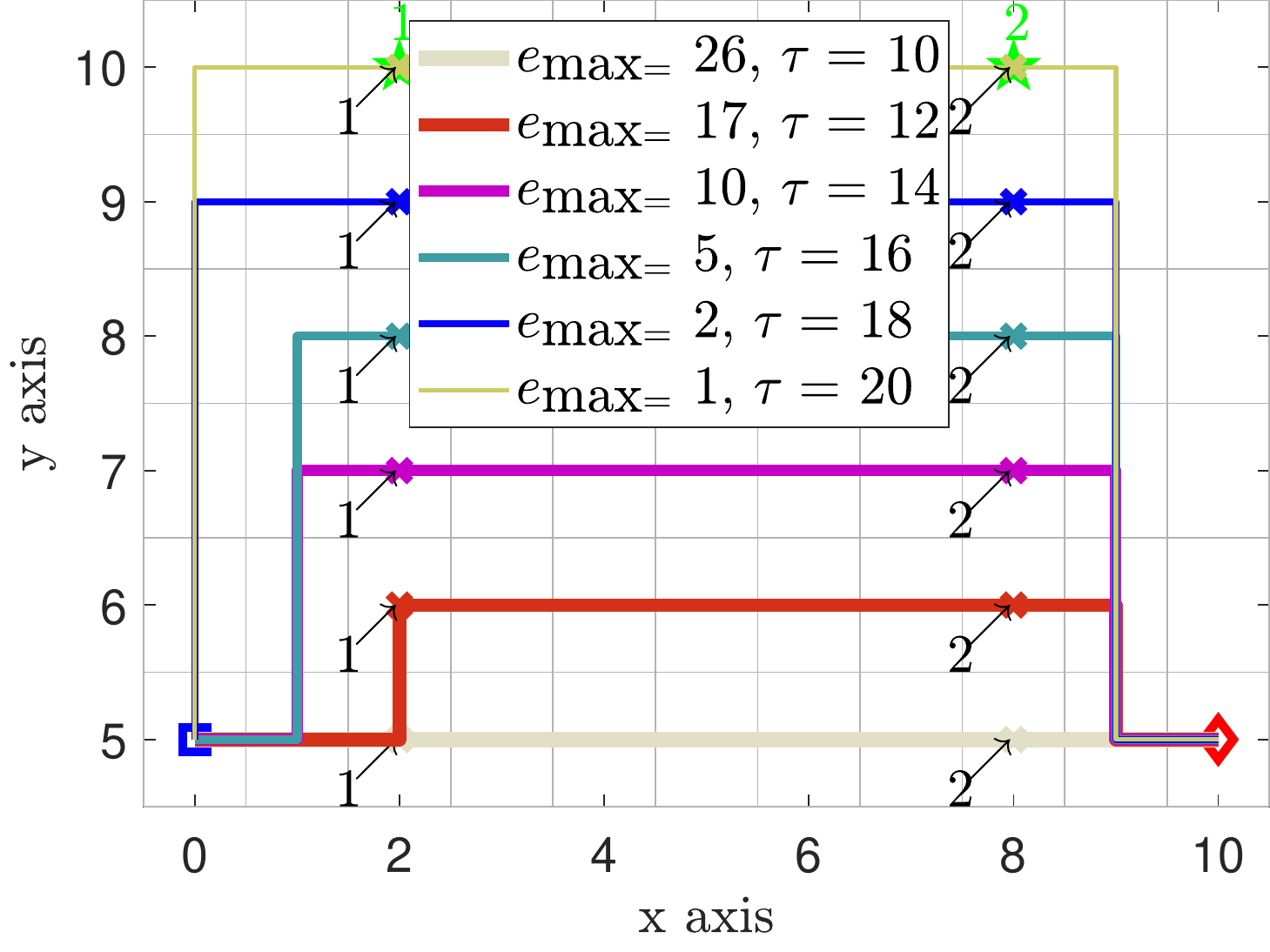}
    \caption{\small Setup for scenario 2.}
    \label{fig:setup_2}
    \end{subfigure}%
    ~
    \begin{subfigure}[b]{0.25\textwidth}
    \centering
    \includegraphics[width=\columnwidth]{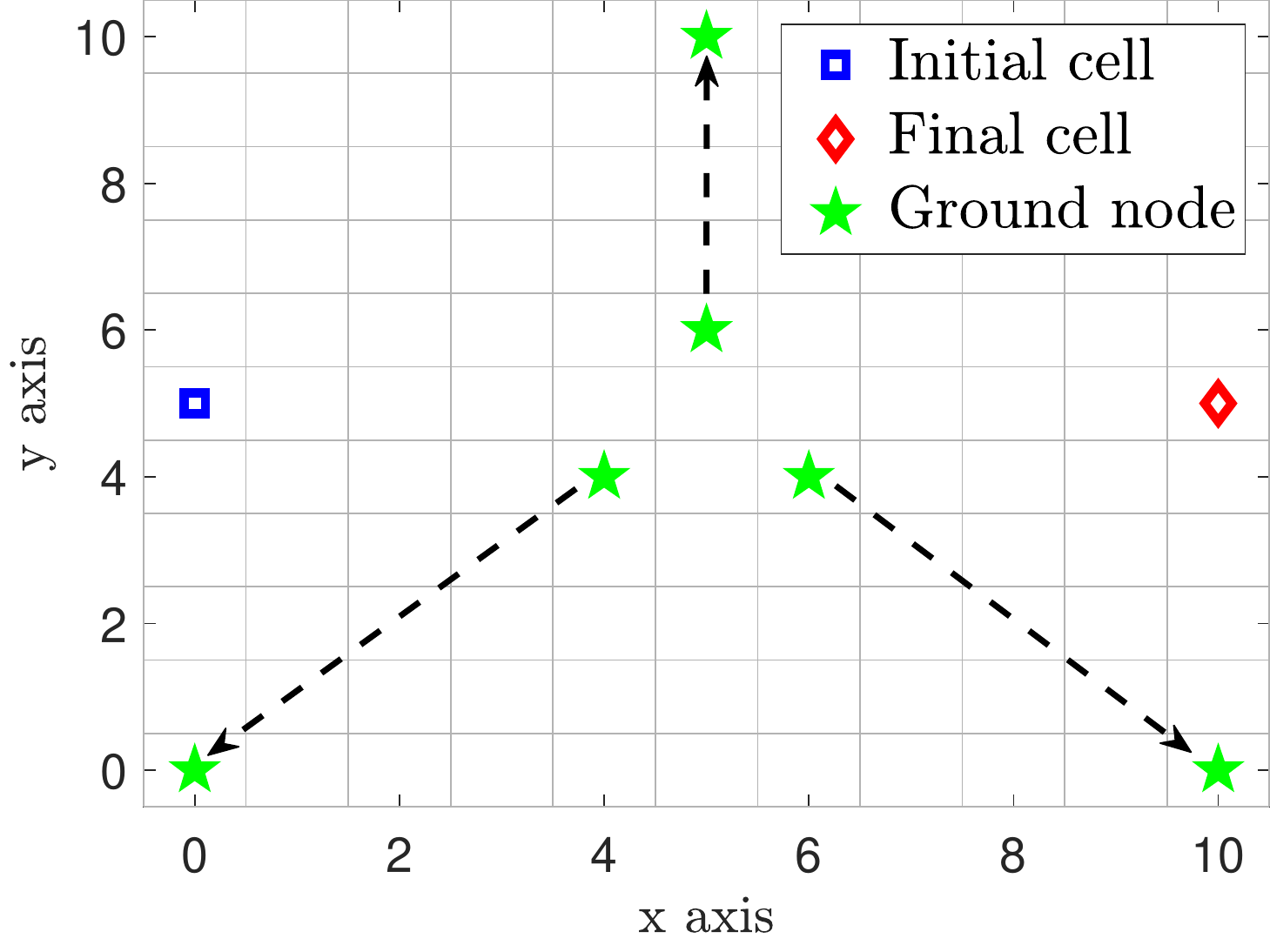}
    \caption{\small Setup for scenarios 3, 4, and 5.}
    \label{fig:setup_345}
    \end{subfigure}
    \caption{\small The initial and final cells, and locations of ground nodes for the five studied scenarios.}
    \label{fig:setups}
    \vspace{-5mm}
\end{figure*}\vspace{-3mm}
\section{Simulation Results}\label{sec:results}
For our simulations, we consider an area in between the following coordinates: $[-50,-50], [-50,1050],[1050,1050],$ and $[1050,-50]$ meters. We discretize this area into cells of dimensions 100 meters by 100 meters where the index of every cell is the coordinate of the cell center divided by 100. For instance, the cell in between $[450,450], [450,550],[550,550],$ and $[550,450]$ meters is called $(5,5)$ since the center of this cell is $[500,500]$ meters. Thus, we will have 11 cells in both the x and y directions. In addition, we consider $ B = 1$ MHz, $ S = 20$ Mbits, $ \sigma^2 = -100$ dBm, $ h = 100$ meters, $L_{u}^{\rm i} = [0,5]$, $L_{u}^{\rm f} = [10,5]$, $\frac{E_{\textrm{max},m}}{e_{\textrm{max},m}} = 1$ mJ, $A_{\textrm{max},m} = 50$ and $\lambda_{m} = \frac{1}{M}, \forall m \in \mathcal{M}$. In order to train the UAV, we consider the ANN architecture in \cite{mnih2015human} with no convolutional neural networks and only one FC layer with 200 hidden nodes. We use the Tensorflow-Agents library \cite{TFAgents} for designing the environment, policy, and costs. In addition, we use a single NVIDIA P100 GPU and 20 Gigabits of memory to train the UAV. Note that the reported numbers are derived by averaging the sum-AoI per process over 1000 episodes. 
% We also assume that UAV's initial and final locations are at cells $[0,5]$ and $[10,5]$, respectively. In addition, we consider $A_{\textrm{max},m} = 50$ for all observed processes, which have equal importance weights. We evaluate the impact of battery size, time constraint, and the location and spatial density of the ground nodes on the sum-AoI per process (we use \emph{per process} since we consider equal weights). 
%The learning rate of the UAV depends on the scale of the state space, and, hence, in all of our simulations, we stop the training after the sum-AoI per process converges to a constant value.
\subsection{Convergence Analysis}
To analyze the convergence of the proposed deep RL algorithm, we illustrate our setup for simulation scenario 1 in Fig. \ref{fig:setup_1}. In this scenario, we have only one ground node which is located at $(5,5), (5,6), \dots,$ or $(5,10)$. Also, we have $\tau = 10$ which is the required number of time slots needed to move directly from the initial cell to the final cell. In addition, we have $ e_{\textrm{max}} = 26 $, which is the number of energy quanta required to transmit packet from the ground node at cell $ (5,10) $ (the furthest cell) to the UAV at cell $(5,5)$.

\begin{figure}[t!]
    \centering
    \includegraphics[width=0.55\columnwidth]{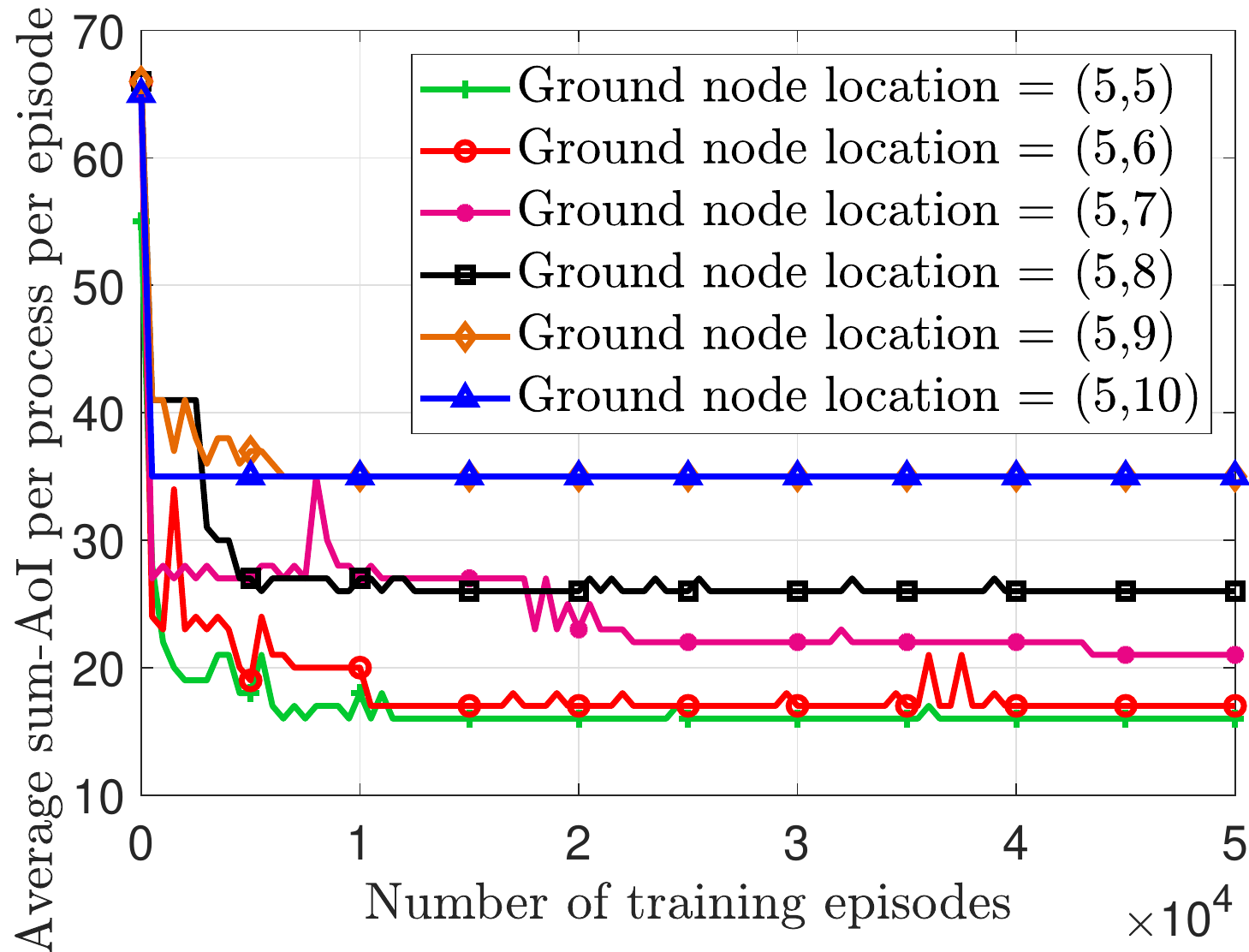}
    \caption{\small Convergence of the average sum-AoI in scenario 1.}
    \label{fig:Scenario_1_convergence}
\end{figure}

Fig. \ref{fig:Scenario_1_convergence} shows the convergence of the average sum-AoI per process (recall that the importance weights are equal) after 50,000 training episodes. We can observe that the average sum-AoI per process is smaller for closer ground nodes to the straight line between the initial and the final cells. This is due to the fact that the UAV has to move in a straight line from the initial cell towards the final cell and, thus, does not have enough energy to update the status of far away ground nodes while the closer nodes can be updated several times.
\vspace{-3mm}
\subsection{Trajectory Optimization}
To show the effective trajectory optimization and scheduling of the UAV, in Fig. \ref{fig:setup_2}, we consider simulation scenario 2 in which there are two ground nodes located at $(2,10)$ and $(8,10)$. Then, we choose $e_{\textrm{max}}$ and $ \tau $ such that the UAV can receive only one packet from any of ground nodes when it can be as close as possible to ground nodes. For instance, when $\tau = 16 $, the UAV will have 6 more time slots than moving straight from initial cell to final cell. Thus, the UAV can use the extra 6 slots to go three slots to the North, update ground nodes and then come back to the straight line. In this case, we choose $e_{\textrm{max}} = 5$ which is the required energy to transmit a packet to a UAV that is located two cells away. Fig. \ref{fig:setup_2} shows that the proposed deep RL algorithm can optimally find the best path and scheduling strategy. The cross marker is the UAV's location at which it receives a status update packet from the ground node with an index next to the cross marker.
\vspace{-6mm}
\subsection{Effects of System Parameters on the Minimum Sum-AoI}
To compare the performance of our proposed deep RL algorithm with other policies, in
Fig. \ref{fig:setup_345}, we set up three scenarios. In scenario 3, we consider three ground nodes located at $(5,10)$, $(0,0)$, and $ (0,10) $, where $ \tau = 100$ and $e_{\textrm{max}}$ varies between $2^0$ and $2^{10}$ and are equal for all of the ground nodes. In scenario 4, the locations of ground nodes are the same as scenario 3 while $e_{\textrm{max}} = 100$ for all of the ground nodes and $\tau $ varies between $10$ and $100$. Scenario 5 studies the effect of the spatial density of ground nodes at the outcome of the optimal policy. To this end, in scenario 5, we have $e_{\textrm{max}} = 100$ for all ground nodes and $\tau =100 $. In addition, the location of ground nodes varies from $(4,4)$, $(5,6)$, and $ (6,4) $ (the most dense case) to $(0,0)$, $(5,10)$, and $ (0,10) $ (the least dense case). We compare the deep RL policy with two baseline policies: 1) a distance-based policy which updates the status of the closest ground node if the distance is less than 2 cells and moves closer to the ground node with the maximum current AoI value, and 2) a random walk policy which randomly chooses a ground node to update its status while moving randomly in all directions. The distance-based policy is heuristically a good policy since it requires less energy for status update and tries to move closer to ground nodes with higher AoI to update their status. On the other hand, the random walk policy always explores all of the actions, thus, may find some actions that are not trivial but will result in smaller average AoI.

\begin{figure*}[t!]
    \centering
    \begin{subfigure}{0.25\textwidth}
        \centering
        \includegraphics[width=\textwidth]{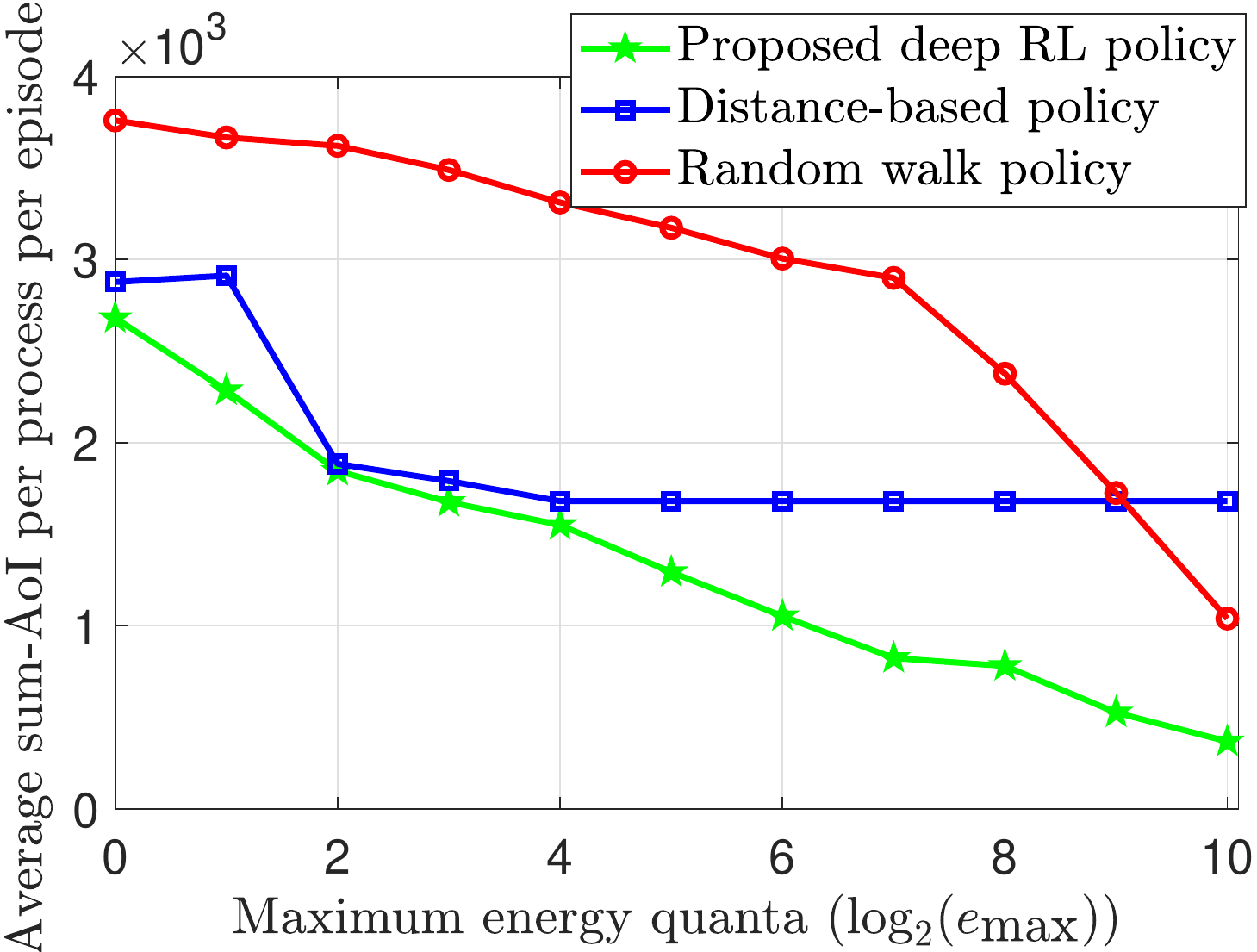}
        \caption{}
        \label{fig:Scenario_3}
    \end{subfigure} 
    ~
    \begin{subfigure}{0.25\textwidth}
        \centering
        \includegraphics[width=\textwidth]{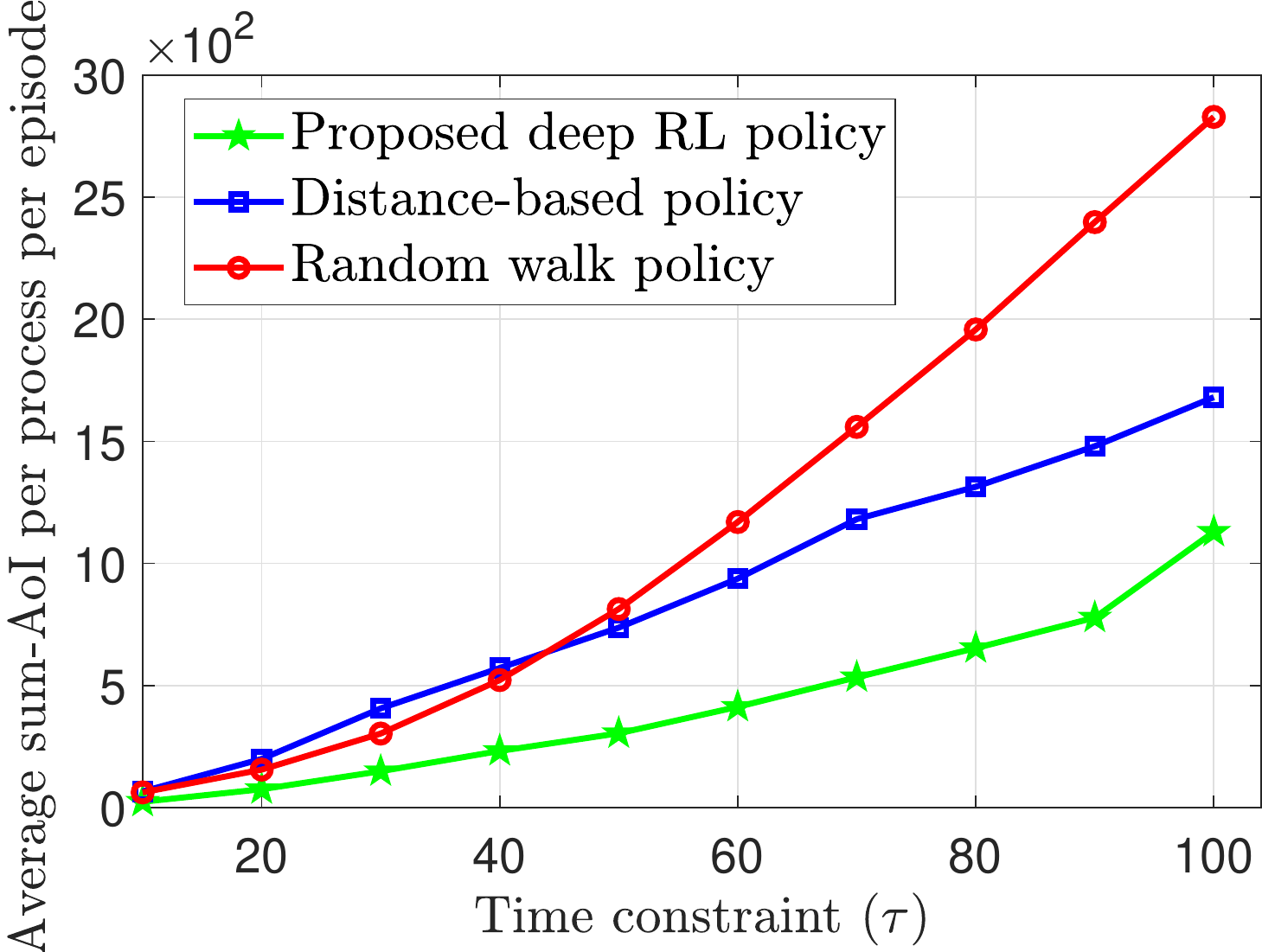}
        \caption{}
        \label{fig:Scenario_4}
    \end{subfigure}
    ~
    \begin{subfigure}{0.25\textwidth}
        \centering
        \includegraphics[width=\textwidth]{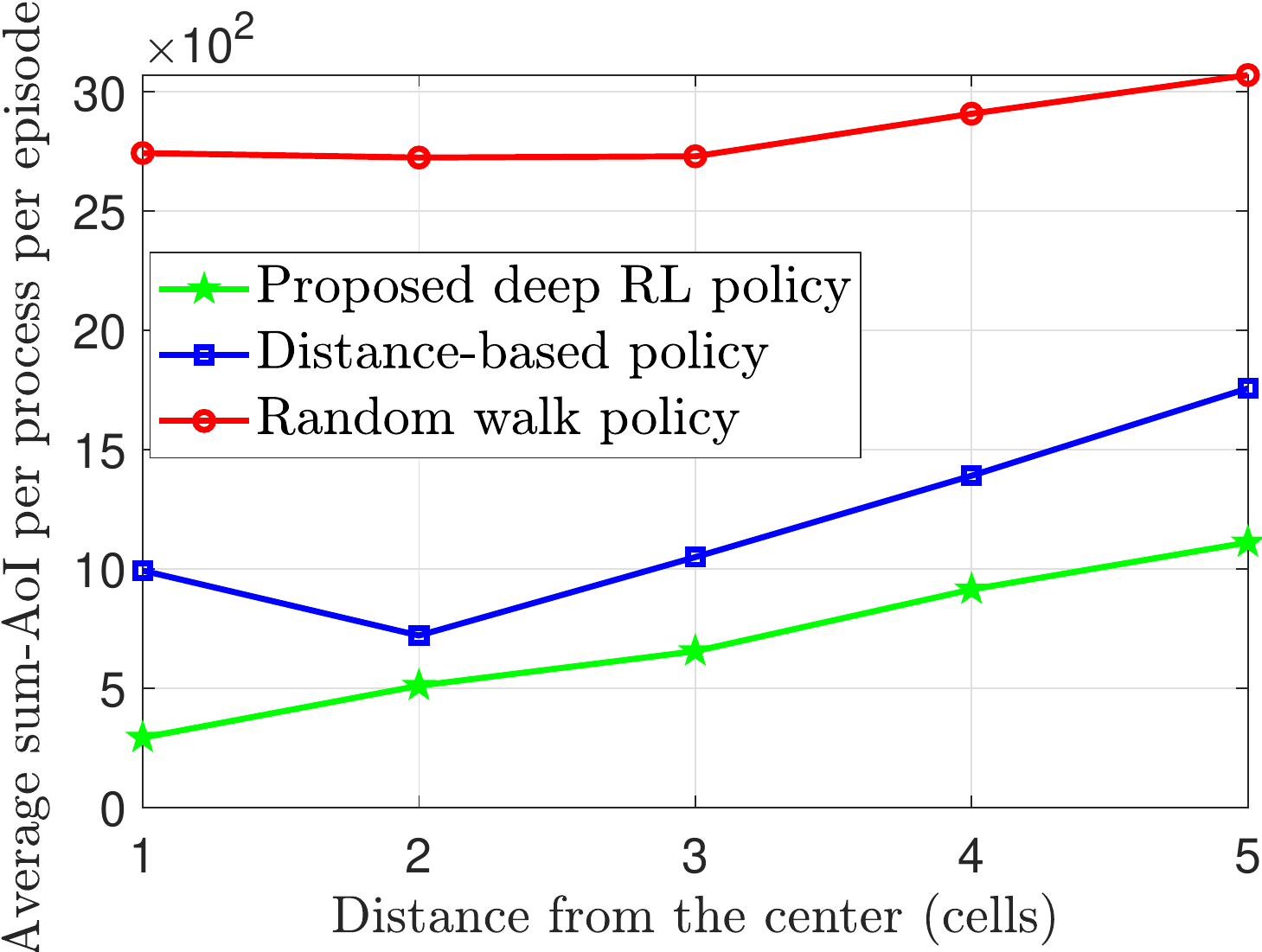}
        \caption{}
        \label{fig:Scenario_5}
    \end{subfigure}
    \caption{\small Average sum-AoI per process vs.: (a) maximum energy quanta, (b) UAV's time constraint, and (c) spatial density of ground nodes.}
    \label{fig:params}
    \vspace{-5mm}
\end{figure*}

Fig. \ref{fig:Scenario_3} shows the effect of $e_{\textrm{max}}$ on the sum-AoI per process in scenario 3. we can observe that a higher $e_{\textrm{max}}$ results in lower average sum-AoI per process since the ground nodes can be updated more frequently and from larger distances. In addition, we can see that our proposed deep RL policy outperforms the other baseline policies since it takes into account the available energy quanta, AoI, the time constraint, and location of the UAV while the other policies are only distance-based or completely random. Fig. \ref{fig:Scenario_3} demonstrates that the distance-based policy is more effective than the random walk policy for smaller $e_{\textrm{max}}$. However, for larger $e_{\textrm{max}}$, the random walk policy is more effective since it can explore more state-action pairs and can update ground nodes from a farther distance. On the other hand, the distance-based policy stays constant after $e_{\textrm{max}} = 2 ^ 5 $ because the agent has to satisfy the time constraint, thus, an increase in $e_{\textrm{max}}$ will not be effective.

Fig. \ref{fig:Scenario_4} shows the results for scenario 4 in which the effect of $\tau$ on the sum-AoI per process is studied. Two key points can be deduced from Fig. \ref{fig:Scenario_4}: 1) the proposed deep RL policy results in approximately $50\%$ and $75\%$ smaller average sum-AoI respectively compared to the distance-based policy and the random walk policy, and 2) for the time constraint smaller than 50, the random walk policy is more effective. However, for larger time constraints, we can see that the distance-based policy has enough time to get closer to ground nodes to update their status, thus, can outperform the random walk policy.

Fig. \ref{fig:Scenario_5} shows the effect of the spatial density of ground nodes on the sum-AoI per process in scenario 5. We can see from Fig. \ref{fig:Scenario_5} that the proposed deep RL policy has a lower average sum-AoI per process compared to the baseline policies. Fig. \ref{fig:Scenario_5} also shows that as the spatial density of ground nodes reduces, i.e., the distance between the ground nodes increases, the average sum-AoI per process increases. This is because, for larger distances, the UAV does not have enough time to get closer to the ground nodes, thus, it has to receive packets from farther distances and less frequently.
\vspace{-3mm}
\section{Conclusion}\label{sec:con}
In this paper, we have investigated the problem of minimizing the weighted sum-AoI for a UAV-assisted wireless network in which a UAV collects status update packets from energy-constrained ground nodes. We have shown that the proposed age-optimal policy can jointly optimize the UAV's flight trajectory as well as scheduling of status update packets from the ground nodes. We have then developed a deep RL algorithm to characterize the age-optimal policy that overcomes the curse of dimenstionality of the original MDP. We have shown that the deep RL algorithm significantly outperforms baseline policies such as the distance-based and random walk policies, in terms of the achievable sum-AoI per process. Numerical results have demonstrated that the achievable sum-AoI per process by the proposed algorithm is monotonically increasing (monotonically decreasing) with the time constraint of the UAV and spatial density of the ground nodes (the battery sizes of the ground nodes).
 %while jointly optimizing the UAV's flight trajectory as well as scheduling of status update packets.
% the achievable sum-AoI per process by the proposed deep RL algorithm, is: i) monotonically decreasing with respect to the battery sizes of the ground nodes, and ii) monotonically increasing with respect to the time constraint of the UAV as well as the spatial density of ground nodes.
 \vspace{-3mm}
\bibliographystyle{IEEEtran}
\bibliography{Globecome_AoI_v0.2}

\begin{thebibliography}{10}
\providecommand{\url}[1]{#1}
\csname url@samestyle\endcsname
\providecommand{\newblock}{\relax}
\providecommand{\bibinfo}[2]{#2}
\providecommand{\BIBentrySTDinterwordspacing}{\spaceskip=0pt\relax}
\providecommand{\BIBentryALTinterwordstretchfactor}{4}
\providecommand{\BIBentryALTinterwordspacing}{\spaceskip=\fontdimen2\font plus
\BIBentryALTinterwordstretchfactor\fontdimen3\font minus
  \fontdimen4\font\relax}
\providecommand{\BIBforeignlanguage}[2]{{%
\expandafter\ifx\csname l@#1\endcsname\relax
\typeout{** WARNING: IEEEtran.bst: No hyphenation pattern has been}%
\typeout{** loaded for the language `#1'. Using the pattern for}%
\typeout{** the default language instead.}%
\else
\language=\csname l@#1\endcsname
\fi
#2}}
\providecommand{\BIBdecl}{\relax}
\BIBdecl

\bibitem{mozaffari2019tutorial}
M.~Mozaffari, W.~Saad, M.~Bennis, Y.-H. Nam, and M.~Debbah, ``A tutorial on
  uavs for wireless networks: Applications, challenges, and open problems,''
  \emph{IEEE Commun. Surveys \& Tutorials}, 2019.

\bibitem{challita2019machine}
U.~Challita, A.~Ferdowsi, M.~Chen, and W.~Saad, ``Machine learning for wireless
  connectivity and security of cellular-connected {UAVs},'' \emph{IEEE Wireless
  Commun.}, vol.~26, no.~1, pp. 28--35, Feb. 2019.

\bibitem{azari2016joint}
M.~M. Azari, F.~Rosas, K.-C. Chen, and S.~Pollin, ``Joint sum-rate and power
  gain analysis of an aerial base station,'' in \emph{Proc. of IEEE Global
  Commun. Workshops (GC Wkshps), Washington, DC}, Dec. 2016.

\bibitem{bor2016efficient}
R.~I. Bor-Yaliniz, A.~El-Keyi, and H.~Yanikomeroglu, ``Efficient {3-D}
  placement of an aerial base station in next generation cellular networks,''
  in \emph{Proc. of IEEE Intl. Conf. on Commun. (ICC), Kuala Lumpur}, May 2016.

\bibitem{chetlur2017downlink}
V.~V. Chetlur and H.~S. Dhillon, ``Downlink coverage analysis for a finite
  {3-D} wireless network of unmanned aerial vehicles,'' \emph{IEEE Trans. on
  Commun.}, vol.~65, no.~10, pp. 4543--4558, Oct. 2017.

\bibitem{alzenad20173}
M.~Alzenad, A.~El-Keyi, F.~Lagum, and H.~Yanikomeroglu, ``{3-D} placement of an
  unmanned aerial vehicle base station {(UAV-BS)} for energy-efficient maximal
  coverage,'' \emph{IEEE Wireless Commun. Letters}, vol.~6, no.~4, pp.
  434--437, Aug. 2017.

\bibitem{zeng2016throughput}
Y.~Zeng, R.~Zhang, and T.~J. Lim, ``Throughput maximization for uav-enabled
  mobile relaying systems,'' \emph{IEEE Trans. on Commun.}, vol.~64, no.~12,
  pp. 4983--4996, Dec. 2016.

\bibitem{li2018placement}
P.~Li and J.~Xu, ``Placement optimization for uav-enabled wireless networks
  with multi-hop backhauls,'' \emph{Journal of Commun. and Information
  Networks}, vol.~3, no.~4, pp. 64--73, Dec. 2018.

\bibitem{xie2018throughput}
L.~Xie, J.~Xu, and R.~Zhang, ``Throughput maximization for uav-enabled wireless
  powered communication networks,'' \emph{IEEE Internet of Things Journal},
  2018.

\bibitem{monwar2018optimized}
M.~Monwar, O.~Semiari, and W.~Saad, ``Optimized path planning for inspection by
  unmanned aerial vehicles swarm with energy constraints,'' in \emph{Proc. of
  IEEE Global Commun. Conf. (GLOBECOM), Abu Dhabi, United Arab Emirates}, Dec.
  2018.

\bibitem{abd2018role}
M.~A. Abd-Elmagid, N.~Pappas, and H.~S. Dhillon, ``On the role of
  age-of-information in internet of things,'' {\em IEEE Commun. Magazine}, to
  appear. Available online: arxiv.org/abs/1812.08286.

\bibitem{kaul2012real}
S.~Kaul, R.~Yates, and M.~Gruteser, ``Real-time status: How often should one
  update?'' in \emph{Proc. of IEEE Conf. on Computer Commun., Orlando, FL},
  March 2012.

\bibitem{kosta2017age_mono}
A.~Kosta, N.~Pappas, and V.~Angelakis, ``Age of information: A new concept,
  metric, and tool,'' \emph{Foundations and Trends in Networking}, vol.~12,
  no.~3, pp. 162--259, Nov. 2017.

\bibitem{ABedewy2016}
A.~M. Bedewy, Y.~Sun, and N.~B. Shroff, ``Optimizing data freshness,
  throughput, and delay in multi-server information-update systems,'' in
  \emph{Proc. of IEEE Intl. Symposium on Information Theory, Barcelona}, July
  2016.

\bibitem{sun2017update}
Y.~Sun, E.~Uysal-Biyikoglu, R.~D. Yates, C.~E. Koksal, and N.~B. Shroff,
  ``Update or wait: How to keep your data fresh,'' \emph{IEEE Trans. on Info.
  Theory}, vol.~63, no.~11, pp. 7492--7508, Nov. 2017.

\bibitem{8648525}
E.~T. Ceran, D.~G{\"u}nd{\"u}z, and A.~Gy{\"o}rgy, ``Average age of information
  with hybrid {ARQ} under a resource constraint,'' \emph{IEEE Trans. on
  Wireless Commun.}, vol.~18, no.~3, pp. 1900--1913, March 2019.

\bibitem{113882}
I.~Kadota, A.~Sinha, and E.~Modiano, ``Optimizing age of information in
  wireless networks with throughput constraints,'' in \emph{Proc. of IEEE Conf.
  on Computer Commun., Honolulu, HI}, April 2018.

\bibitem{AbdElmagid2019Globecom_a}
M.~A. Abd-Elmagid, H.~S. Dhillon, and N.~Pappas, ``Online age-minimal sampling
  policy for {RF}-powered {IoT} networks,'' in \emph{Proc. of IEEE Global
  Commun. Conf. (GLOBECOM)}, Dec. 2019.

\bibitem{talak2018optimizing}
R.~Talak, S.~Karaman, and E.~Modiano, ``Optimizing age of information in
  wireless networks with perfect channel state information,'' in \emph{Proc. of
  Intl. Symposium on Modeling and Optimization in Mobile, Ad Hoc and Wireless
  Networks, Shanghai}, May 2018.

\bibitem{zhou2018joint}
B.~Zhou and W.~Saad, ``Joint status sampling and updating for minimizing age of
  information in the {Internet} of {Things},'' 2018, available online:
  arXiv.org/abs/1807.04356.

\bibitem{abdel2018ultra}
M.~K. Abdel-Aziz, C.-F. Liu, S.~Samarakoon, M.~Bennis, and W.~Saad,
  ``Ultra-reliable low-latency vehicular networks: Taming the age of
  information tail,'' in \emph{Proc. of IEEE Global Commun. Conf. (GLOBECOM),
  Abu Dhabi, United Arab Emirates}, Dec. 2018.

\bibitem{abd2018average}
M.~A. Abd-Elmagid and H.~S. Dhillon, ``Average peak age-of-information
  minimization in {UAV}-assisted {IoT} networks,'' \emph{IEEE Trans. on Veh.
  Technology}, vol.~68, no.~2, pp. 2003--2008, Feb. 2019.

\bibitem{8406973}
J.~{Liu}, X.~{Wang}, B.~{Bai}, and H.~{Dai}, ``Age-optimal trajectory planning
  for {UAV}-assisted data collection,'' in \emph{Proc. of IEEE Conf. on
  Computer Commun. Workshops (INFOCOM WKSHPS), Honolulu, HI}, April 2018.

\bibitem{powell2007approximate}
W.~B. Powell, \emph{Approximate Dynamic Programming: Solving the curses of
  dimensionality}.\hskip 1em plus 0.5em minus 0.4em\relax John Wiley \& Sons,
  2007, vol. 703.

\bibitem{mnih2015human}
V.~Mnih, K.~Kavukcuoglu, D.~Silver, A.~A. Rusu, J.~Veness, M.~G. Bellemare,
  A.~Graves, M.~Riedmiller, A.~K. Fidjeland, G.~Ostrovski, S.~Petersen,
  C.~Beattie, A.~Sadik, I.~Antonoglou, H.~King, D.~Kumaran, D.~Wierstra,
  S.~Legg, and D.~Hassabis, ``Human-level control through deep reinforcement
  learning,'' \emph{Nature}, vol. 518, no. 7540, p. 529, 2015.

\bibitem{TFAgents}
S.~Guadarrama, A.~Korattikara, O.~Ramirez, P.~Castro, S.~F. Ethan~Holly, E.~G.
  Ke~Wang, C.~Harris, V.~Vanhoucke, and E.~Brevdo, ``{TF-Agents}: A library for
  reinforcement learning in tensorflow,''
  \url{https://github.com/tensorflow/agents}, 2018.

\end{thebibliography}
\end{document}